\documentclass[aip,amsmath,amssymb,reprint,]{revtex4-1}

\usepackage{graphicx}
\usepackage{dcolumn}
\usepackage{bm}
\usepackage[utf8]{inputenc}
\usepackage[T1]{fontenc}
\usepackage{mathptmx}
\usepackage{hyper ref}

\usepackage[separate-uncertainty=true]{siunitx}

\usepackage{color}
\newcommand{\red}[1]{\textcolor{black}{#1}}

\newcommand{\new}[1]{\textcolor{black}{#1}}

\begin{document}

\preprint{}

\title{Electromagnetic induction imaging with atomic magnetometers: unlocking the low-conductivity regime}

\author{Luca Marmugi}
\email{l.marmugi@ucl.ac.uk}
\author{Cameron Deans}
\author{Ferruccio Renzoni}
\affiliation{Department of Physics and Astronomy, University College London, Gower Street, London, WC1E 6BT, UK}

\date{\today}

\begin{abstract}
Electromagnetic induction imaging with atomic magnetometers has disclosed unprecedented domains for imaging, from security screening to material characterization. However, applications to low-conductivity specimens -- most notably for biomedical imaging -- require sensitivity, stability, and tunability only speculated thus far. Here, we demonstrate contactless and non-invasive imaging down to \SI{50}{\siemens\per\metre} using a \SI{50}{\femto\tesla\per\sqrt{\hertz}} $^{87}$Rb radio-frequency atomic magnetometer operating in an unshielded environment and near room temperature. Two-dimensional images of test objects are obtained with a near-resonant imaging approach, which reduces the phase noise by a factor 172, with projected sensitivity of \SI{1}{\siemens\per\metre}. Our results, an improvement of more than three orders of magnitude on previous imaging demonstrations, push electromagnetic imaging with atomic magnetometers to regions of interest for semiconductors, insulators, and biological tissues.

\vskip 20pt
\begin{center}
This is a preprint version of the article appeared in Applied Physics Letters:\\
L. Marmugi, C. Deans, F. Renzoni, Appl. Phys. Lett. {\bf 115}, 083503 (2019) DOI: \href{https://doi.org/10.1063/1.5116811}{10.1063/1.5116811}.
\end{center}

\end{abstract}

%%%%%% Measured phase noise at 0 Hz: 1.67°. At 750 Hz: 0.077°. Ratio: 22.
%%%%%% Projected maximum phase noise (5 nT) at 0 Hz: 6.347°. At 750 Hz: 0.037°. Ratio: 172.

\maketitle

%\section{Introduction}
Imaging is ubiquitous throughout science, technology, and everyday	life. However, some areas remain \red{not directly accessible}. This is because of the lack of dedicated instrumentation, or the limited sensitivity of established technologies.  This is the case of conductivity mapping of biological tissues and organs, which is precluded to currently deployed diagnostic devices. Contactless and non-invasive access to such information would provide insight and diagnostic tools for several conditions, whose clinical phenotype features modifications of the conductivity of the involved tissues. Examples include degenerative neurological and muscular diseases, fibrosis, skin healing, liver, lung and mammary glands tumors\cite{tumours}, hemorrhages, edemas, and cardiac fibrillations. Conditions such as atrial fibrillation have enormous human and economic costs due to non-ideal diagnostics and limited access to the fundamental pathogenic mechanisms\cite{scirep2016}.

A possible solution to the problem of identifying suitable imaging devices is magnetic induction tomography \cite{griffiths2001}, which has been proposed for many applications in the biomedical field\cite{griffiths1999, merwa2004, zolgharni2010, rabbit}.  In this approach, an oscillating magnetic field induces eddy currents, whose density depends on the local properties of the specimens, and a magnetometer measures the secondary field produced by them. However, the limited sensitivity and tunability of the magnetic sensors in use -- most often pick-up coils -- have prevented suitable imaging performance. Detection of low-conductivity objects ($\approx\SI{1}{\siemens\per\m}$) has only been achieved with conventional sensors for very large volumes (e.g., $\gg\SI{2000}{\cm\cubed}$) \cite{griffiths2007residual}. 

Electromagnetic induction imaging with radio-frequency atomic magnetometers\cite{apl2016, wickenbrock2016} has the potential to overcome such limitations. It has been successfully demonstrated in a number of applications, including localization and identification of non-metallic targets\cite{prl2018}, material characterization\cite{prl2018, apl2018, witold2018, witoldnew}, and security screening\cite{opex2017, ao2018}.  Thus far, imaging has been limited to large conductivities ($\geq\SI{e5}{\siemens\per\metre}$). Recently, high-conductivity imaging has been also obtained with nitrogen-vacancy (NV) center magnetometers \cite{wickenbrock2018}, where the proximity to the specimen reachable with a nano-sized sensor can compensate for the current lack of sensitivity. However, none of these demonstrations has fully met the requirements for practical uses in conductivity mapping of biological tissues. Namely: i) high sensitivity for measuring the small response of non-conductive specimens; ii) broad tunability for controlling the penetration and compensating for the decrease in conductivity; \red{iii) high-phase stability in unshielded environments for ensuring reliable imaging;} and \red{iv) large dynamic range (i.e. the ability to simultaneously image large differences in conductivity) to image specimens with large variations in conductivity, without any detrimental saturation effect.}

In this Letter, we address the above points, demonstrating 2D electromagnetic induction imaging of \SI{6.25}{\cm\cubed} specimens down to \SI{50}{\siemens\per\metre}.  Images are obtained with a \SI{50}{\femto\tesla\per\sqrt{\hertz}}, broadly tunable $^{87}$Rb radio-frequency (RF) atomic magnetometer\cite{budker2007, sensitivity, sensitivity2, sensitivity3}, operating in a magnetically noisy environment. Specimens are imaged in real time, in the form of a phase map, directly proportional to the local conductivity. Phase stability better than $\SI{0.03}{\degree}$  across the imaging area over several hours is obtained by biasing the RF atomic magnetometer to near-resonant operation. In this way, the impact of residual magnetic noise is reduced by two orders of magnitude. Simultaneous imaging of samples spanning three decades in conductivities is also demonstrated. Our results represent a 1460-fold improvement in conductivity over previous works performing electromagnetic imaging with atomic magnetometers \cite{prl2018} -- extending its domain to low-conductivity specimens. This moves applications to biomedical imaging closer \cite{scirep2016}.

%\section{Experimental setup and protocol}
\begin{figure*}[htpb]
\includegraphics[width=0.85\linewidth]{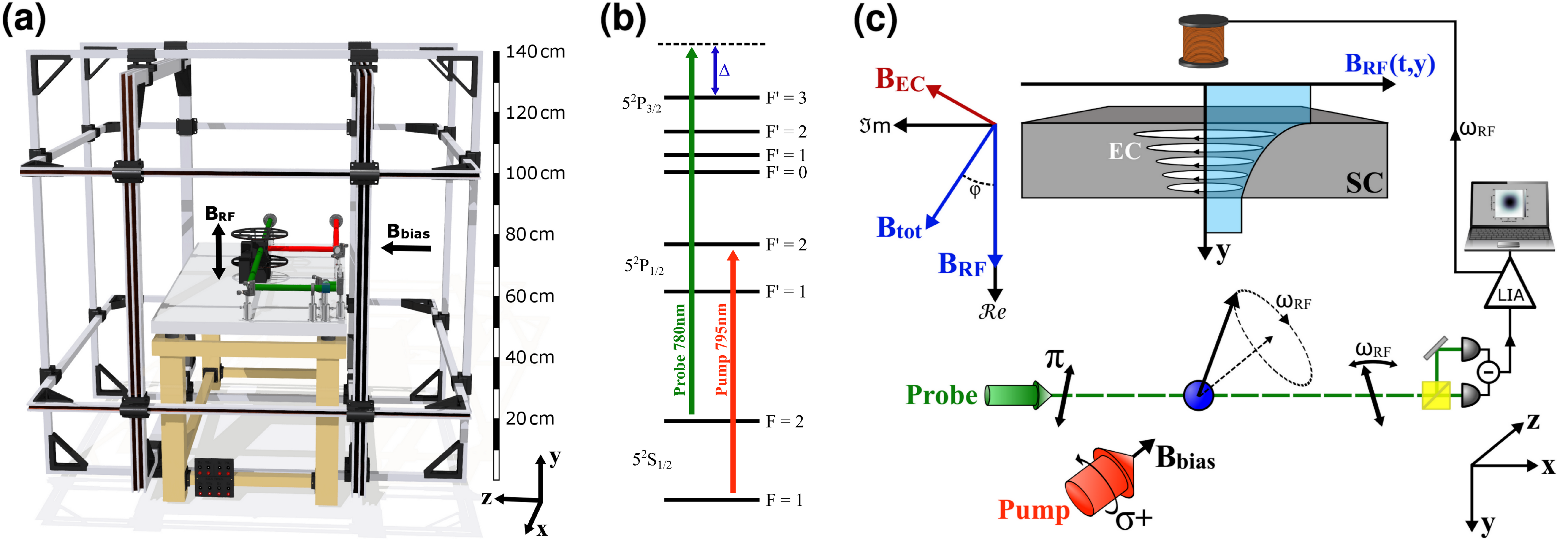}
\caption{{\bf (a)} Sketch -- to scale -- of the \SI{50}{\femto\tesla\per\sqrt{\hertz}} RF atomic magnetometer for electromagnetic induction imaging of low-conductivity specimens (laser systems not shown). The red and green lines mark the optical paths of the pump and the probe laser beams, respectively. The dark arrows indicate the orientation of the bias field (B$_{\text{bias}}$) and of the driving field (B$_{\text{RF}}$). {\bf (b)} $^{87}$Rb level  diagram and tuning of lasers. {\bf (c)} Principle of the imaging technique. LIA: lock-in amplifier. SC: semiconductor sample. EC: eddy currents. Details are in the main text.}\label{fig:amemi}
\end{figure*}

A sketch of the setup is shown in Fig.~\ref{fig:amemi}(a). The core of the system is a cubic quartz chamber containing isotopically enriched  $^{87}$Rb and  \SI{40}{\text{Torr}} of N$_{2}$ as buffer gas. The cube has a side of \SI{25}{\milli\meter}. The vapor cell is housed in a 3D printed Nylon case, hosting an active temperature control system (\SI{4.6}{\watt} capacity). The vapor is maintained at \SI{45}{\celsius}. The sensor is held in place by a polylactide (PLA) mount, secured to a marble table equipped with Sorbothane vibrational isolators.%, placed around \SI{70}{\centi\metre} away from the center of sensor.
 
A three-axes Helmholtz and a three-axes quadrupole coil sets ensure homogeneity of the magnetic field at the sensor's position by cancelling stray fields and gradients, respectively. A DC bias field (B$_{\text{bias}}$, along $\hat{z}$) is actively stabilized by a proportional-integrative-derivative (PID) loop referenced to a fluxgate in close proximity to the atomic magnetometer. The fluxgate continuously monitors the background magnetic field. Its output is used as the error signal of a feedback loop\cite{rsi2018, witold2018}. The magnetic field compensation system has a response band between DC and \SI{1}{\kilo\hertz}, imposed by the fluxgate.

Atomic spins are optically pumped by a circularly polarized laser detuned by $+\SI{80}{\mega\hertz}$ with respect to the D$_{1}$ line ($5^{2}$S$_{1/2}|$F=1$\rangle \to5^{2}$P$_{1/2}|$F$^{\prime}=2\rangle$), at \SI{795}{\nano\meter}. Tuning of the magnetometer is obtained with B$_{\text{bias}}$, collinear to a pump beam along $+\hat{z}$ (See Fig.~\ref{fig:amemi}(a)). Optical pumping accumulates atoms in the $|F=2$, m$_{\text{F}}=+2\rangle$ state. B$_{\text{RF}}$ (generated by a single ferrite-core coil of \SI{7.8}{\milli\meter} diameter, and oscillating along $\pm\hat{y}$) coherently drives atomic Zeeman coherences at $\omega_{\text{RF}}$. Precessing spins are probed by a second laser, linearly polarized and tuned to the D$_{2}$ line (\SI{1.35}{\giga\hertz} to the blue side of the $5^{2}$S$_{1/2}|$F=2$\rangle \to5^{2}$P$_{3/2}|$F$^{\prime}=3\rangle$ transition) at \SI{780}{\nano\meter} (Fig.~\ref{fig:amemi}(b)). The probe beam propagates along $+\hat{x}$. Detection of Larmor precession is obtained through Faraday rotation of the probe beam's polarization plane. The rotation is measured by a polarimeter connected to a lock-in amplifier, referenced to $\omega_{\text{RF}}$. Throughout this work, the amplifier's time constant T$_{\text{C}}$ is set to \SI{500}{\milli\second}. Operation of the RF atomic magnetometer in the range between \SI{100}{\kilo\hertz} and \SI{2}{\mega\hertz} -- the upper limit currently imposed by the acquisition electronics -- is obtained.

%\subsection{EMI images creation and display}
Unlike conventional atomic magnetometry, which relies on the measurement of spontaneously generated magnetic signals \cite{sienaheart, romalis, danesi, oxfordneuron}, in our approach B$_{\text{RF}}$(t, y) induces eddy currents in the samples, with a density exponentially decaying along $\hat{y}$ (Fig.~\ref{fig:amemi}(c)) as determined by the skin effect\cite{vander2006rf}, and \red{a magnetic signature proportional to $\omega_{RF}^{2}$, as already reported\cite{griffiths2007residual}.} Specimens are placed in between the RF coil and the atomic magnetometer, \SI{30}{\milli\metre} above it. However, this arrangement is not a requirement for electromagnetic induction imaging with atomic magnetometers \cite{apl2018,witold2018,witoldnew}. Samples are supported by a motorized PLA and Nylon support (step size \SI{2}{\milli\metre}). Eddy currents produce a secondary field (B$_{\text{EC}}$) opposing B$_{\text{RF}}$ and phase-lagged with respect to it. B$_{\text{EC}}$ contains information on the conductivity $\sigma$, permittivity $\varepsilon$, permeability $\mu$, and the geometry of the samples\cite{peyton2017, peyton2018}. Such information is retrieved by monitoring the influence of B$_{\text{EC}}$ on the atomic spins' motion. Data is stored in position-referenced matrices. Each pixel is the result of a single measurement. Datasets are then smoothed with a nearest-neighbor Gaussian filter (radius \SI{4}{\milli\metre}), and plotted in color-coded 2D graphs. The color map is scaled between the maximum and the minimum of each single image, unless otherwise indicated. In the following, the symbol $\Delta \Phi$ is used to convey the idea of a variation from the sensor's equilibrium (i.e.\ no eddy currents) induced by the sample and the related B$_{\text{EC}}$. However, no background subtraction is performed.

%\section{Experimental results}
\begin{figure*}[htbp]
\includegraphics[width=0.99\linewidth]{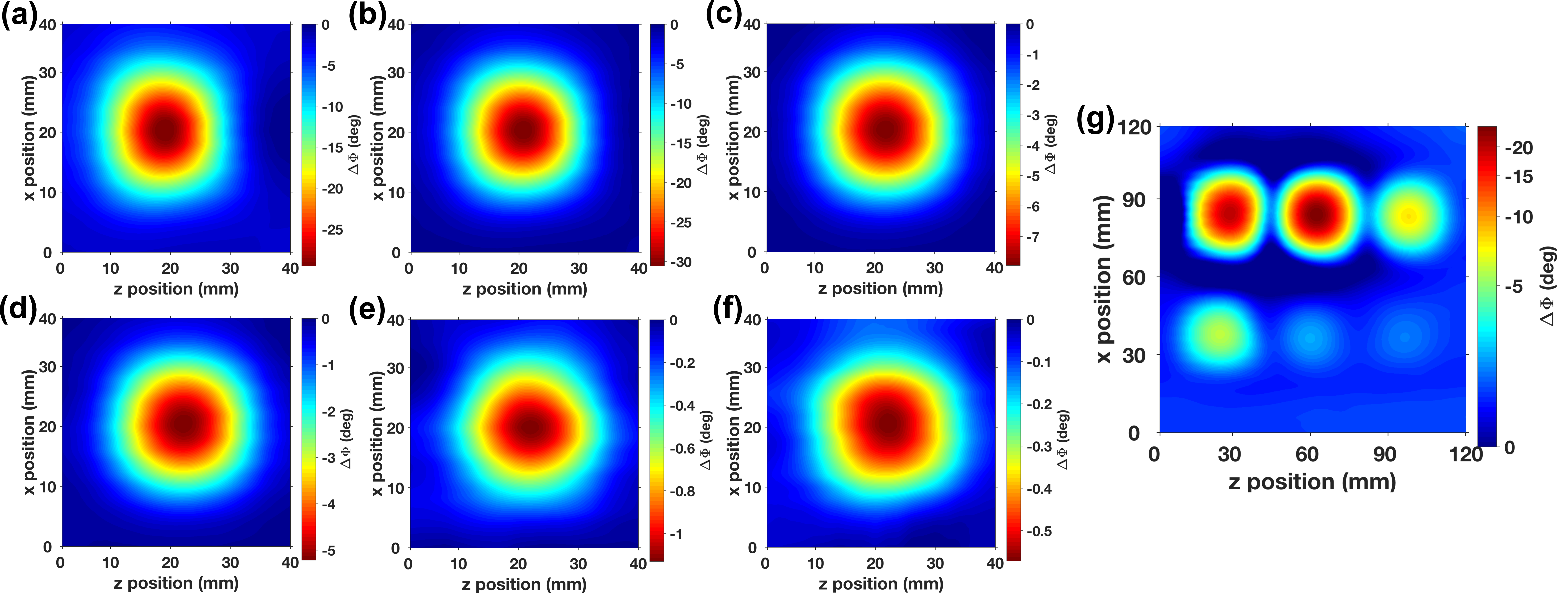}
\caption{Imaging of low-conductivity samples with an RF atomic magnetometer operating at \SI{2}{\MHz}. $\Delta \Phi$ scans of the doped Si specimens.  {\bf (a)} $\sigma=\SI{1e4}{\siemens\per\metre}$. {\bf (b)}  $\sigma=\SI{5e3}{\siemens\per\metre}$. {\bf (c)} $\sigma=\SI{1e3}{\siemens\per\metre}$. {\bf (d)} $\sigma=\SI{5e2}{\siemens\per\metre}$. {\bf (e)} $\sigma=\SI{1e2}{\siemens\per\metre}$. {\bf (f)}  $\sigma=\SI{5e1}{\siemens\per\metre}$. \red{ {\bf (g)} Simultaneous imaging at \SI{2}{\mega\hertz} of specimens with conductivities spanning $[\SI{1e4}{\siemens\per\metre}, \SI{5e1}{\siemens\per\metre}]$. Conductivity decreases from top left to right bottom. Samples are not in electrical contact.}}
\label{fig:images}
\end{figure*}

Figures~\ref{fig:images}(a)-(f) show phase maps ($\Delta \Phi$), obtained at \SI{2}{\MHz}, spanning conductivities from $\sigma=\SI{1e4}{\siemens\per\metre}$ to \SI{5e1}{\siemens\per\metre}. \red{This frequency was chosen as the upper limit of $\omega_{\text{RF}}$, imposed by the acquisition electronics.} Samples are $25\times25\times\SI{10}{\milli\metre\cubed}$ Si blocks with decreasing n-doping concentrations (P, from \SI{5e18}{\per\centi\meter\cubed} to \SI{2e15}{\per\centi\meter\cubed}). Phase maps are presented as they are directly related to $\sigma$. Furthermore, variations in phase cannot be produced by shielding or shadowing of the sensor.

In all cases, samples are clearly imaged with satisfactory definition. The \SI{5e1}{\siemens\per\metre} sample surpasses the previous record by 1460 times \cite{prl2018}. The maximum recorded $\Delta \Phi$ decreases with the samples' conductivity and hence their dopant levels -- with an approximately linear relationship between $\Delta \Phi$ and $\omega_{\text{RF}}\sigma$. Correspondingly, the phase variation at a given frequency becomes smaller, as $\sigma$ decreases, unless $\omega_{\text{RF}}$ is suitably increased. At \SI{2}{\mega\hertz} the skin depth $\delta(\omega_{\text{RF}})$ is smaller than the sample thickness for the most conductive specimens. This reduced penetration into the bulk explains why the maximum $\Delta\Phi$ for the \SI{1e4}{\siemens\per\metre} sample (Fig.~\ref{fig:images}(a)) is smaller than that of the \SI{5e3}{\siemens\per\metre} one (Fig.~\ref{fig:images}(b)). 

\begin{figure}[htbp]
\includegraphics[width=\linewidth]{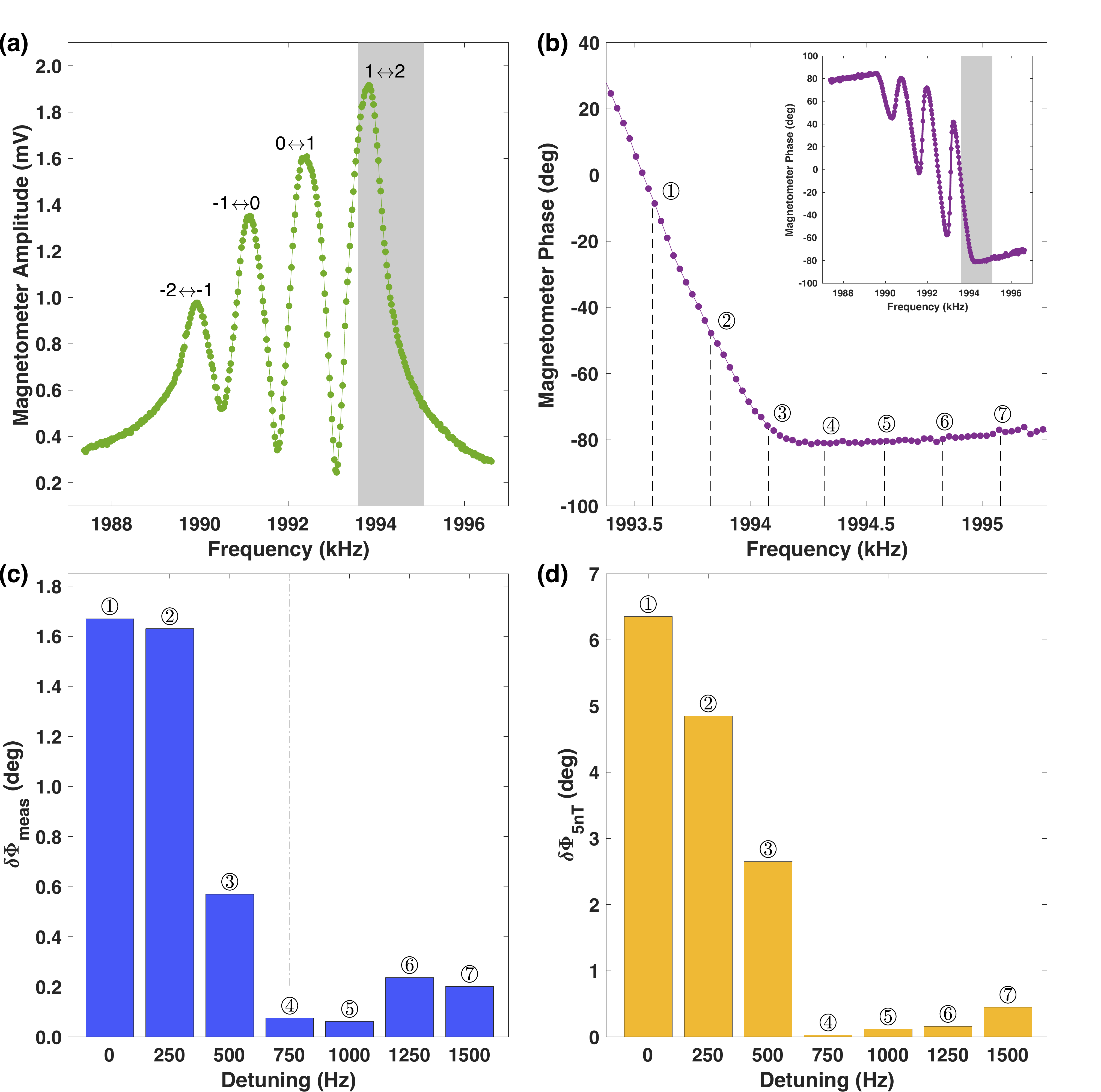}
\caption{Near-resonant imaging. {\bf (a)} Magnetometer amplitude response near \SI{2}{\mega\hertz}. The non-linear Zeeman effect causes the RF resonance to split in four components, \red{corresponding to individual transitions $|F=2, m_{F}\rangle \leftrightarrow |F=2, m_{F}\pm1\rangle$, as indicated by the respective labels.} The shaded area highlights the region explored for near-resonant operation. {\bf (b)} Magnetometer phase profile and operating points for near-resonant imaging. Inset: the entire phase response highlighting the near-resonant region. {\bf (c)}  Example of the measured phase noise ($\delta\Phi_{\text{meas}}$) across the imaging area \new{(the total phase noise in an image without specimens)} versus detuning. {\bf (d)} Calculated phase noise ($\delta\Phi_{5\text{nT}}$) produced by \SI{5}{\nano\tesla} magnetic field noise, as measured across the imaging area. \red{The circled numbers mark the position with respect to the resonance curve in panel (b).} The vertical dotted lines mark the configuration used throughout this work.}\label{fig:nearresonant} 
\end{figure}

\red{Imaging single specimens does not provide any indication of the response in the presence of abrupt and large variations in conductivity. This is another essential requirement for biomedical imaging and navigation: for example, cancerous tissues can exhibit increases in conductivity up to 500 times\cite{tumours}. To demonstrate the large dynamic range and the conductivity sensitivity of our system, a simultaneous image at \SI{2}{\mega\hertz} of the six samples of different conductivities is shown in Fig.~\ref{fig:images}(g). This covers a change in conductivity of a factor 200. The low-conductivity samples exhibit a decreasing contrast but are clearly visible -- with signal levels consistent with the previous images.}

Images shown in \red{Figs.~\ref{fig:images}(a)-(g)} were only possible after reducing the noise-induced phase variations. These hamper the imaging process: for imaging low-conductivity, phase stability is as critical as the sensitivity and the tunability of the RF atomic magnetometer. Residual magnetic noise shifts the magnetometer frequency and perturbs the recorded phase, imposing a limit on the smallest detectable phase change and therefore on the imaging performance. Magnetic noise is greatly reduced by operating within magnetically shielded environments, however this is not suitable for practical uses of electromagnetically induction imaging. Alternatively, self-tuning optimization could significantly improve the performance of active compensation systems for atomic magnetometers \cite{sienaoptimisation}.

Instead, we introduce near-resonant \new{induction imaging with} RF atomic magnetometers, as illustrated in Fig.~\ref{fig:nearresonant}.  Such a solution can be extended to other magnetometers -- independently of the performance of the field stabilization, if applicable -- and to any resonant, phase-sensitive system characterized by residual phase noise limiting its performance. 

With this technique, the RF atomic magnetometer is biased to operate in quasi-resonant conditions (Fig.~\ref{fig:nearresonant}(a)), where the gradient of the phase response is the smallest (Fig.~\ref{fig:nearresonant}(b)). This minimizes the impact of magnetic field noise on the system's output whilst maintaining the sensitive detection of phase changes $\Delta\Phi$ arising from B$_\text{EC}$.  To quantify the performance improvement provided by near-resonant imaging, the phase noise is recorded across the imaging area ($\delta \Phi_{\text{meas}}$) at $\omega_{\text{RF}}=\SI{2}{\mega \hertz}$, as shown in Fig.~\ref{fig:nearresonant}(c). In other words, $\delta \Phi_{\text{meas}}$ represents the total phase noise recorded whilst imaging in absence of a specimen. The optimum working point is where $\delta \Phi_{\text{meas}}$ is minimal (i.e.\ highest phase stability). Estimates of the phase noise are obtained from the residual magnetic field noise after active stabilization -- \SI{5}{\nano\tesla}, regardless the operational frequency. The intrinsic phase noise limit ($\delta\Phi_{\text{5nT}}$) is calculated from this value, and used for identifying the best operational conditions (Fig.~\ref{fig:nearresonant}(d)). The near-resonant detuning point is dependent on the operation frequency as must be chosen accordingly. For example, at $\omega_{\text{RF}} = \SI{200}{\kHz}$ the detuning is $+\SI{375}{\Hz}$. In the case of Fig.~\ref{fig:nearresonant}, in resonant conditions, the minimum detectable conductivity is \SI{160}{\siemens\per\metre}, as a result of the phase noise. Near-resonant operation at \SI{2}{\mega\hertz}, with a detuning of $+\SI{750}{\hertz}$, reduces the phase noise by a factor 172, thus bringing the residual noise phase to $\delta\Phi_{5\text{nT}}=\SI{0.03}{\degree}$. This projects imaging to the \SI{1}{\siemens\per\metre} level, fully matching the requirements for imaging biological tissues\cite{scirep2016} ($\leq\SI{10}{\siemens\per\metre}$).

\begin{figure}[htbp]
\includegraphics[width=\linewidth]{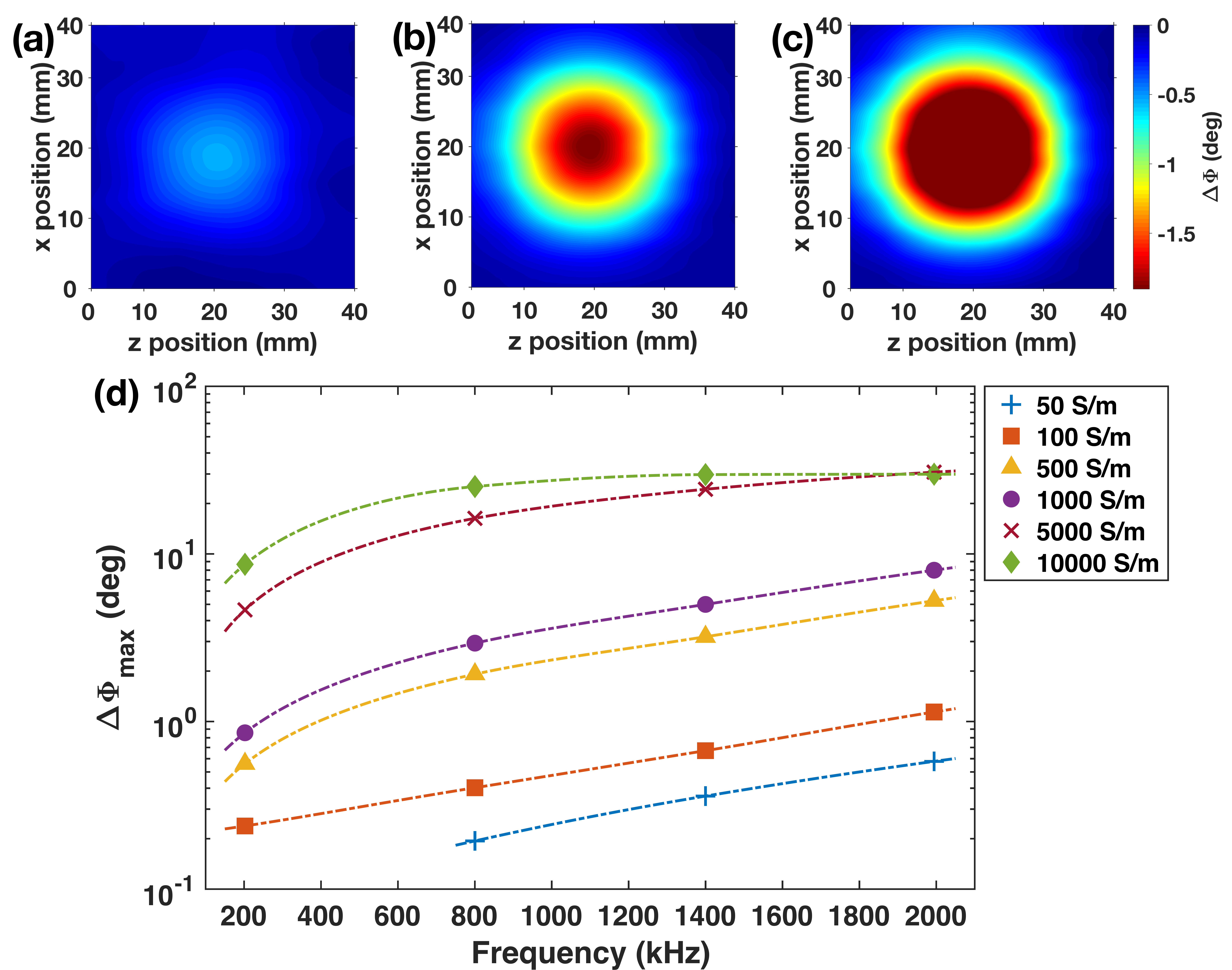}
\caption{\red{Imaging tunability. \textbf{(a)-(c)} Images of the \SI{5e2}{\siemens\per\metre} sample at \SI{200}{\kilo\hertz}, \SI{800}{\kilo\hertz}, and \SI{1.4}{\mega\hertz}. \textbf{(d)} Maximum phase change across an image ($\Delta \Phi_{\text{max}}$) for each of the six doped Si samples as a function of frequency. Dashed lines are the best interpolating curves (piecewise cubic Hermite interpolating polynomials) for each dataset.}}
\label{fig:tunability}
\end{figure}

\red{Figure \ref{fig:tunability} demonstrates the tunability of the imaging system: in panels (a)-(c), we show the \SI{5e2}{\siemens\per\metre} sample imaged at \SI{200}{\kilo\hertz}, \SI{800}{\kilo\hertz}, and \SI{1.4}{\mega\hertz}. Otherwise, the conditions are the same of Fig.~\ref{fig:images}. The corresponding ratios between the samples' thickness ($t=\SI{10}{\milli\metre}$) and the skin depth $\delta_{\omega_{\text{RF}}}$ are: 0.20, 0.40, and 0.53, respectively (0.63 when $\omega_{\text{RF}}=\SI{2}{\mega\hertz}$ -- Fig.~\ref{fig:images}(d)). For a penetration depth larger than the object's thickness, a linear dependence of the phase on the frequency is expected\cite{griffiths2007residual}, as anticipated:}

\begin{equation}
\red{\Delta\Phi \approx \omega_{\text{RF}}\mu_{0}\sigma~,}\label{eqn:sigma}
\end{equation}

\red{where $\mu_{0}$ is the magnetic permeability of free space. This can be observed in the case of the low-conductivity samples, for which this condition is satisfied -- See lower part of Fig.~\ref{fig:tunability}(d). For the higher conductivities, deviations from Eq.~\ref{eqn:sigma} can be noticed in the upper part of Fig.~\ref{fig:tunability}(d), and are attributed to the skin effect. This regulates the generation of eddy currents, and thus the volume contributing to the imaging signal. At high frequency, the skin depth for the high-conductivity specimens is smaller than their thickness (e.g., for the \SI{1e4}{\siemens\per\metre} sample, $t/\delta_{\omega\text{RF}}\approx2.8$ at \SI{2}{\mega\hertz}). Thus, the effective volume is reduced, and a lower increase of $\Delta \Phi_{\text{max}}$ with frequency is observed. The results of Fig.~\ref{fig:tunability} also demonstrate that our system is capable of resolving variations in conductivity $\leq 2$, thus covering a wide range of cases of interest for biomedical imaging\cite{deansspie}.}

%\section{Conclusions}
In conclusion, we have demonstrated contactless and non-invasive electromagnetic induction imaging of low-conductivity specimens with an ultra-sensitive (\SI{50}{\femto\tesla\per\sqrt{\hertz}}) and broadly tunable (\SI{100}{\kilo\hertz}-\SI{2}{\mega\hertz}) RF atomic magnetometer biased to operate in near-resonant conditions. Imaging was obtained in unshielded environments, near room temperature. Images of low-conductivity test samples -- \SI{6.25}{\cm\cubed} doped semiconductors measuring between \SI{1e4}{\siemens\per\metre} and \SI{5e1}{\siemens\per\metre} -- represent a three orders of magnitude reduction in the lowest-conductivity material imaged with an atomic magnetometer, and a substantial decrease in volume with respect to previous results with low-conductivity samples. A large dynamic range was also observed, with the largest span of conductivities ($[\SI{1e4}{\siemens\per\metre}, \SI{5e1}{\siemens\per\metre}]$) simultaneously imaged, to date. Near-resonant operation allowed the reduction of the phase-noise in unshielded environment to \SI{0.03}{\degree} in operational conditions. Projected performance of our system indicate sensitivity to \SI{1}{\siemens\per\metre} in unshielded environment, \red{with a sensitivity to changes in conductivity smaller than a factor 2}, thus matching the requirements for long-sought applications in biomedical imaging. \red{Here, variations of conductivity between $10^{2}$  and 10 are expected, on a scale of several millimeters\cite{scirep2016, tumours}. In light of the results presented in this Letter, this performance appears achievable in the very short term. Dedicated optimisation and validation for specific conditions would be then necessary.}

\begin{acknowledgments}
This work was supported by the UK Quantum Technology Hub in Sensing and Metrology, Engineering and Physical Sciences Research Council (EPSRC) (EP/M013294/1).
\end{acknowledgments}

\nocite{*}

\end{document}